\documentclass[conference]{IEEEtran}
\IEEEoverridecommandlockouts

\usepackage{cite}
\usepackage{amsmath,amssymb,amsfonts}
\usepackage{algorithmic}
\usepackage{graphicx}
\usepackage{textcomp}
\usepackage{multirow}
\usepackage{array}
\usepackage{xcolor}
\usepackage{color}
\usepackage{adjustbox}
\usepackage{bm}

\newcommand{\balpha}{\bm{\alpha}}

\def\BibTeX{{\rm B\kern-.05em{\sc i\kern-.025em b}\kern-.08em
 T\kern-.1667em\lower.7ex\hbox{E}\kern-.125emX}}
\begin{document}

\title{Configurable Multilingual ASR with Speech Summary Representations\\
}

\author{\IEEEauthorblockN{Harrison Zhu, Ivan Fung, Yingke Zhu, Lahiru Samarakoon }
\IEEEauthorblockA{
\textit{Fano}\\
Hong Kong SAR, China \\
$\{\text{harrison.zhu, ivan, yingke.zhu, lahiru}\}$@fano.ai}}

\maketitle

\begin{abstract}
  Approximately half of the world's population is multilingual, making multilingual ASR (MASR) essential. Deploying multiple monolingual models is challenging when the ground-truth language is unknown in advance. This motivates research efforts on configurable multilingual MASR models that can be prompted manually or adapted automatically to recognise specific languages. In this paper, we present the Configurable MASR model with Summary Vector (csvMASR), a novel architecture designed to enhance configurability. Our approach leverages adapters and introduces speech summary vector representations, inspired by conversational summary representations in speech diarization, to combine outputs from language-specific components at the utterance level. We also incorporate an auxiliary language classification loss to enhance configurability. Using data from 7 languages in the Multilingual Librispeech (MLS) dataset, csvMASR outperforms existing MASR models and reduces the word error rate (WER) from 10.33\% to 9.95\% when compared with the baseline. Additionally, csvMASR demonstrates superior performance in language classification and prompting tasks.
\end{abstract}


\section{Introduction}
It is believed that approximately half of the world's population is multilingual \cite{garcia_bilingual_2011}, making multilingual speech recognition essential. Moreover, we often do not know the ground-truth language in advance and only know that the language spoken is from a range of languages, making it difficult to deploy a specific monolingual model. This has motivated the study of multilingual ASR (MASR) models as an alternative option. However, naively training a MASR model with pooled multilingual training data usually leads to inferior performance compared to monolingual models \cite{zhou_configurable_2021,kwon_mole_2023,sun_building_2023}, which can be due to language confusion and bias towards data-rich languages \cite{kwon_mole_2023}.

To improve MASR models, one can study configurable MASR models. Configurable MASR models can both manually be prompted to a subset of desired languages (e.g. based on prior knowledge or business requirements) and automatically adapt to the ground-truth language. Configurability is crucial in achieving improved performances \cite{zhou_configurable_2021,sun_building_2023} as well as being the focus of many recent works \cite{zhou_configurable_2021,sun_building_2023,hu_mixture--expert_2023,kwon_mole_2023,wang_language-routing_2023}.

A popular configurable MASR technique, that we refer to as LIDConcat, brings substantial performance improvements \cite{li_multi-dialect_2018,waters_leveraging_2019} by leveraging the language ID (LID) vector (a binary-valued vector that indicates the presence of languages). LIDConcat concatenates the ground-truth LIDs with the input features during training and inference by passing the resulting vector into the encoder \cite{li_multi-dialect_2018,waters_leveraging_2019}. However, this may be problematic when the ground-truth LID is only known up to a range of languages, as incorrect LIDs will also then need to be inputted and may cause the features to become contaminated with incorrect embeddings. When the model is trained in a multihot LID setting, the LIDConcat approach also gives very little improvement to the model's performance \cite{zhou_configurable_2021}. 


Another direction is to incorporate language-specific components into the attention layers \cite{aditya_attention-guided_2024} or at the feed-forward layers through the use of mixtures-of-experts (MoE; \cite{zhou_configurable_2021,hu_mixture--expert_2023,kwon_mole_2023,wang_language-routing_2023}). The language-specific components of the MASR models are crucial to ensuring that the model can adapt to the ground-truth LID, given a prompt containing it. Intuitively, the language-specific components can learn useful language-specific features that are part of more distinct subspaces.

In order to only activate the desired language-specific components to avoid any contamination, we require weighted interpolation mechanism. With a weighted interpolation mechanism, language-specific weights are computed and used to combine the language-specific components. There are several ways to do this, but the most common approaches are:
\paragraph{Fixed Utterance-Level} An LID-prompted uniform weighted interpolation mechanism on the language-specific experts \cite{zhou_configurable_2021}, where the weights are determined by the input LID vector. However, similar to the LIDConcat approach, this may be problematic as the model can be easily contaminated by incorrect LIDs.

\paragraph{Learnable Framewise} Passing the post-attention-feed-forward features through a classifier to obtain framewise language-specific weights \cite{sun_building_2023,hu_mixture--expert_2023,kwon_mole_2023,wang_language-routing_2023}. This produces a framewise weighted interpolation mechanism that can allow us to determine the language-specific weights at the frame level, which suffers from less contamination by incorrect LIDs. However, since each frame also needs to participate in the language classification task, it will inevitably need to store the utterance-level information. This could be at the expense of the feature representation of localised information, and thus harm the model's language classification and ASR performance. Interestingly, a related phenomenon has been shown in vision transformers \cite{darcet_vision_2024} in the form of artifacts in the attention maps.

To accurately determine the language-specific weights without affecting the frame-level feature representations, we could construct additional utterance-level representations for this purpose. We take inspiration from the learnable conversation summary vector (SV) representations in diarization \cite{broughton_improving_2023,fung_robust_2023,samarakoon_transformer_2023}. The success of the SV has been demonstrated in the speech domain of diarization to better encode utterance-level information and thus improve speaker existence prediction. The SV itself is inspired by the BERT \texttt{[CLS]} token \cite{devlin_bert_2019} for downstream classification tasks e.g. sentiment analysis. In another domain of computer vision, the Vision Transformer (ViT) \cite{dosovitskiy_image_2021} also uses the \texttt{[CLS]} token for image classification. Recently, this has also been studied with the use of registers \cite{darcet_vision_2024} as a means to improve the interpretability and performance of ViTs by using them to store global information. What all the aforementioned works have in common is that the SV can be an excellent representation of the global information of the input data, whether it is audio, text or image. However, the use of the SV for both speech processing and non-downstream tasks has only been explored in the context of speaker diarization \cite{broughton_improving_2023,fung_robust_2023,samarakoon_transformer_2023}.

In this work, we propose a novel configurable MASR model: Configurable MASR model with Summary Vector (\textbf{csvMASR}). Firstly, csvMASR takes advantage of the parameter-efficient use of the feed-forward residual connection as a universal module \cite{zhou_configurable_2021}, and improves this design by redirecting the residual feature from the feed-forward layer as the input to the language-specific adapters \cite{houlsby_parameter-efficient_2019,hou_exploiting_2021,bai_efficient_2024}. Secondly, we introduce speech summary vector representations to determine language-specific weights at the utterance level. This encourages the summary vector to focus on learning the utterance-level representation and relieves the model from having to store language information in each frame. Thirdly, an auxiliary language classification loss is added to enhance configurability. Lastly, on a 7-language setup, csvMASR reduces the WER from 10.33\% to 9.95\% when compared with the baseline MASR model. On language classification tasks, csvMASR also performs achieves up to 16.65\% higher accuracy than the Framewise weighted interpolation model. Lastly, language prompting tasks demonstrate its configurability, with a WER gap of $<1\%$ between 1-hot and all-hot LID decodings.

\section{Methodology}
\subsection{Backbone ASR Model}
The audio is first converted to 80-dimensional log-mel filterbanks. For the backbone ASR model, we have a connectionist temporal classification module (CTC; \cite{graves_connectionist_2006}) and a Transformer decoder \cite{vaswani_attention_2017} that make up the Hybrid CTC-Attention model \cite{watanabe_hybrid_2017}. The encoder is a stack of $N$ Conformer layers \cite{gulati_conformer_2020}. The model can perform both autoregressive (AR) and non-autoregressive (NAR) decoding, using Beam Search and CTC greedy decoding respectively.

\subsection{Parameter-Efficient Adapters}
We denote the output of the feed-forward module after the attention module as $h_0$, and the outputs of the $L$ language-specific experts $h_1,\ldots,h_L$. Following CMM \cite{zhou_configurable_2021}, $h_0$ is a residual connection to the expert outputs $h_1\ldots h_L$ 
\begin{align}
 h = h_0 + \sum_{i=1}^{L} \alpha_i h_i,
\end{align}
where $\alpha_i$ are the language-specific weights (see section~\ref{sec:SV}). This design has been shown to allow $h_0$ to capture the shared information across languages and require fewer language-specific parameters compared to other common MoE designs.

CMM computes the language-specific expert outputs $h_1,\ldots,h_L$ by feeding the output of the attention module to the $L$ linear expert layers. Firstly, we replace the linear experts in CMM with adapter experts \cite{houlsby_parameter-efficient_2019} to make the model even more parameter-efficient. 

Secondly, $h_0$ is instead fed to the $L$ experts to sequentially connect the feed-forward layer to the language-specific components. This allows us to make use of the richer feature representations in $h_0$. Interestingly, this design is more popular in the literature \cite{houlsby_parameter-efficient_2019,hou_exploiting_2021,le_lightweight_2021,fung_robust_2023,sun_building_2023,chronopoulou_language-family_2023}. 

\subsection{Speech Summary Representations}
\label{sec:SV}
Define the language ID (LID) vector $\mathbf{M}\in\{0,1\}^L$ as a binary vector representing the presence of the $L$ languages e.g. $\mathbf{M}=(0,1,1)$ denotes a 3-hot vector with the second and third languages present. The LID vector $\mathbf{M}$ can be prompted to configure the model to switch on and off the language-specific components. During training, we sample $\mathbf{M}$ by keeping the ground-truth LID and randomly insert incorrect LIDs with probability $p\in(0,1)$. This mimics the real-world scenario where the precise ground-truth LID is unknown.

CMM uses a uniform weighted interpolation mechanism $\alpha_i = \mathbf{M}_i / \sum_{l} \mathbf{M}_l$
that is entirely determined by the input LID vector $\mathbf{M}$.  However, this may be problematic as the model may be easily contaminated by incorrect LIDs when $\mathbf{M}$ contains incorrect LIDs. Conversely, framewise weighted interpolation mechanisms \cite{sun_building_2023,hu_mixture--expert_2023,kwon_mole_2023,wang_language-routing_2023} propose to replace the uniform weighted interpolation mechanism with a learnable language classifier $\varphi$ that produces weights at the frame level, but this could also be problematic as the frame-level features may not contain sufficient language information. Moreover, the utterance-level language information may also need to be stored in the frames (drawing an analogy to a similar observation in vision \cite{darcet_vision_2024}), which could degrade the local linguistic information, and thus harm the ASR performance.

Borrowing the successful implementation and application in speech diarization \cite{broughton_improving_2023,fung_robust_2023,samarakoon_transformer_2023}, we propose to append a learnable summary vector $\theta_\text{SV}$ to the feature vector and propagate it through the encoder, skipping all the convolutions but participating in the attention mechanisms. Skipping the convolutions encourages the summary vector to focus on learning the utterance-level representation and could relieve the model from having to store linguistic information in the frames.

Given this utterance-level representation, we pass it to the classifier module $\varphi$ to produce the language-specific weights $\balpha$. Manual LID prompting can be achieved by applying the mask $\mathbf{M}$ within the softmax, which sets the $\alpha$'s for inactivated LIDs to 0 and normalises the remaining $\alpha$'s to sum to 1.

\subsection{Auxiliary Language Classification Loss}
In addition, to enhance the model's configurability, we add the language classification loss to the final loss $\mathcal{L} = (1-\lambda)(\beta\mathcal{L}_\text{CTC} + (1-\beta)\mathcal{L}_\text{Att}) + \lambda\mathcal{L}_\text{lang}$, where $\lambda, \beta\in[0,1]$ are hyperparameters, $\mathcal{L}_\text{Att}$ is the decoder loss as defined in \cite{vaswani_attention_2017} and $\mathcal{L}_\text{CTC}$ is the the CTC loss as defined in \cite{watanabe_hybrid_2017}. The language classification loss $\mathcal{L}_\text{lang}$ is the cross-entropy loss averaged over all the available adapter layers.  

\begin{figure}[!t]
  \centering
  \includegraphics[width=1\columnwidth]{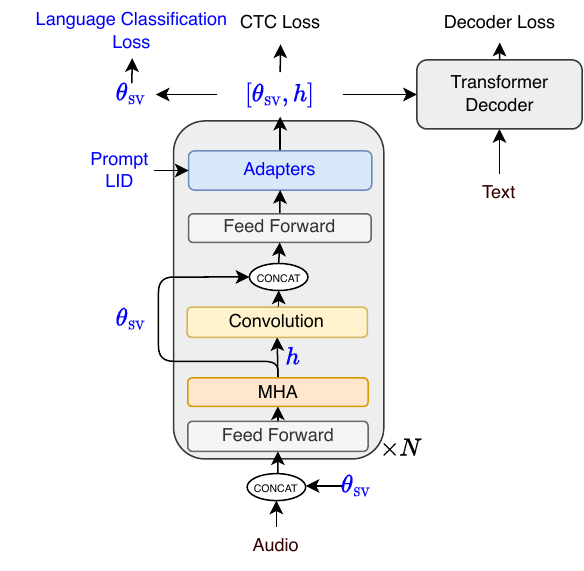}
  \vspace{-9mm}
  \caption{Model architecture of our proposed csvMASR model, with {\color{blue} blue} indicating our novel contributions. The {\color{blue}Adapters} module can also be sparsely inserted into a subset of the $N$ conformer layers for parameter efficiency. The {\color{blue}Prompted LID} refers to the input language ID (LID) vector, which is a binary-valued vector that indicates the presence of languages. The speech summary vector {\color{blue} $\theta_\text{SV}$} skips convolutions and computes the {\color{blue}Language Classification Loss}. Note that {\color{blue} $\theta_\text{SV}$} is also included to compute the CTC Loss and Decoder Loss. }
  \label{fig:csvMASR}
  \vspace{-3mm}
 \end{figure}

Finally, the adapters are placed at the end of the conformer layers and can be sparsely inserted into a subset of the $N$ conformer layers for parameter efficiency. Figure~\ref{fig:csvMASR} shows the overall architecture of our proposed model, Configurable MASR model with Summary Vector (csvMASR), in an end-to-end ASR setting.

\section{Experiments}
\label{sec:exp}
\subsection{Data}
We use the MLS dataset \cite{pratap_mls_2020} and only the 7 non-English languages from the Multilingual Librispeech (MLS) dataset \cite{pratap_mls_2020}. The MLS dataset contains 8 languages: English, Dutch, French, German, Spanish, Italian, Portuguese and Polish. However, we removed English due to its relatively enormous size (44K hours) compared to the rest.

\begin{table}[!t]
  \label{tab:1:Hours}
  \caption{Number of hours for each language in the MLS dataset.}
  \vspace{-3mm}
  \centering
  \begin{adjustbox}{max width=\columnwidth}
    \begin{tabular}{|c|c|c|c|c|c|c|c|c|}
      \hline
      Split & Dutch & French & German & Spanish & Italian & Portuguese & Polish & Aggregate\\
      \hline
      train & 1520 & 1870 & 1034 & 889 & 244 & 137 & 104 & 5798  \\
 val & 12.7 & 14.2  & 9.91 & 9.66 & 5.17 & 2.52 & 2.07 & 66.03  \\
 test & 12.8 & 14.3  & 10.1 & 10.0 & 5.27 & 3.74 & 2.14 & 69.05  \\
    \hline
    \end{tabular}
  \end{adjustbox}
  \vspace{-6mm}
  \end{table}

\subsection{Model Architecture and Training}  
\begin{table*}[!t]
  \label{tab:1:wer_scores}
  \caption{AR/NAR WER for each model. The LID column indicates the language prompt used for inference.}
  \vspace{-3mm}
  \centering
  \begin{adjustbox}{max width=2\columnwidth}
    \begin{tabular}{|c|c|c|c|c|c|c|c|c|c|c|}
    \hline
 Model & Size & LID & Dutch & French & German & Spanish & Italian & Portuguese & Polish & Aggregate\\
    \hline
 Monolingual & 108.65M  & N/A & 14.0/14.7 & 7.2/8.1 & 8.2/9.0 & N/A & N/A & N/A & N/A  & N/A\\
 Multilingual & 109.84M  & N/A & 12.6/13.7 & 7.5/8.6 & 8.2/9.4 & 6.7/7.5 & 13.7/15.4 & 19.0/21.6 & 18.5/20.5 & 10.33/11.56\\
 LIDConcat & 110.36M  & 1-hot & 12.8/13.9 & 7.5/8.6 & 8.2/9.4 & 6.1/6.9 & 13.1/15.0 &18.7/21.5 &16.6/19.3 & 10.13/11.44\\
 & & all-hot & 12.8/13.9 & 7.5/8.6 & 8.2/9.4& 6.2/7.0 &13.1/15.1 &18.8/21.6 &17.0/19.8 & 10.18/11.49\\
 Framewise-3 & 115.37M & 1-hot & 14.2/15.2 & 7.4/8.5 & 8.0/9.0& 5.8/6.5 & 12.7/14.3 &18.4/20.7 & 15.7/18.5 & 10.28/11.45\\
 & & all-hot & 14.2/15.2 & 7.5/8.5 & 8.0/9.0& 5.9/6.6 &12.9/14.6 &20.1/22.6&18.7/21.5 & 10.52/11.70\\
    \hline
 csvMASR-3 (ours) & 115.37M & 1-hot & 12.6/13.7 & 7.4/8.5 & 8.1/9.2 & 5.9/6.7 & 14.0/15.9 & 16.3/18.7 & 17.1/19.2 & \textbf{9.96}/\textbf{11.18}\\
     & & all-hot & 12.6/13.7 & 7.5/8.5 & 8.2/9.3 & 6.0/6.8 & 14.2/16.0 & 16.9/19.5 & 17.3/19.4 & \textbf{10.07}/\textbf{11.29}\\
 csvMASR-4 (ours) & 117.21M & 1-hot &12.6/13.8 & 7.5/8.8 & 8.1/9.6& 6.1/7.0 &13.7/15.7 &16.9/19.6 &14.6/17.0 & \textbf{9.95}/11.40\\
     & & all-hot & 12.7/13.9 & 7.5/8.8 & 8.2/9.6& 6.2/7.1 &13.8/15.9 &18.4/21.0 &14.5/16.8 & \textbf{10.09}/11.50\\
    \hline
    \end{tabular}
  \end{adjustbox}
  \vspace{-5mm}
  \end{table*}

During training, we use $p=0.5$ to sample the multihot LID vectors. We use 12 Conformer layers with CNN kernel size equal to 31. CTC is parameterised by a linear layer with a softmax activation function, and the decoder is parameterised by 6 transformer decoder layers and a linear layer with a softmax activation function. We use 512 attention hidden dimensions, 8 attention heads and 2048 feed-forward network units for both the encoder and decoder. Relative positional encodings are used. Adapters are parameterised by down and up feed-forwards with 256 units in-between. In addition, we only train monolingual baselines for languages with more than 1000 hours of training data to ensure a fair comparison. For the loss weighting, $\lambda=0.5$ and $\beta=0.3$. BPE tokenisation is used with 128 tokens/language. We train the model using the Adam optimizer with a learning rate of 0.0005, for 80 epochs for all MASR models and 100 epochs for monolingual models. The final model is obtained as the averaged model over the best 10 epochs selected using the validation accuracy.

\section{Results}
We compare the csvMASR model with the following models: (1) Monolingual baseline, (2) Multilingual baseline, (3) LIDConcat and (4) Framewise weighted interpolation (Framewise). We experiment with 3 and 4 layers of adapters in csvMASR (csvMASR-3 and csvMASR-4) for comparison. In particular, the adapters are inserted into the 3rd, 6th and 9th layers for csvMASR-3 and the 3rd, 6th, 9th and 12th layers for csvMASR-4. In addition, Framewise-3 is used as a baseline for the framewise weighted interpolation mechanism, with the 3rd, 6th and 9th layers containing adapters.

The word error rate (WER) is used to evaluate the ASR performance in both autoregressive (AR) and non-autoregressive (NAR) settings. Table II shows the WER for each model, with the LID column indicating the inference LID prompt. csvMASR-3 and csvMASR-4 give the best overall WER performance of $9.96\%$ and $9.95\%$ respectively, even without any knowledge of the ground-truth LID (all-hot). This demonstrates that csvMASR models have strong automatic configurability. In addition, the AR performance of csvMASR can be further improved by $0.1\%$ (1-hot) and $0.2\%$ (all-hot) by using the ground-truth LID only (1-hot). However, we see that adding more adapters in csvMASR-4 degrades the NAR performance by $0.22\%$ (1-hot) and $0.21\%$ (all-hot), indicating that adding adapters in the later layers may even harm ASR performance. A hypothesis could be that the model's final layers are already saturated with language information and that the later layers should be reserved for ASR purposes only.

In contrast, there is a small difference of $0.05\%$ between the 1-hot and all-hot WERs for LIDConcat, likely indicating that the LID is effectively ignored by the model. We also tested out the performance of the model without the ground-truth LID and the WER did not change significantly, reinforcing this fact. In contrast, the performance of csvMASR-3 degraded massively when the ground-truth LID was not inputted, indicating that the model relies heavily on the LID information. The performance of Framewise-3 is the worst of all the models, likely due to language confusion issues or the loss of frame-level linguistic information. In addition, the WER gap of $0.24\%$ between 1-hot and all-hot is larger compared to $0.11\%$ of csvMASR-3, indicating a lack of configurability.

Compared to monolingual models, all models perform similarly on the French and German test sets. However, the monolingual AR performance of $14.0\%$ is worse on Dutch than $12.6\%/12.7\%/12.8\%$ for all models except Framewise-3. This is likely due to insufficient output units or that multilingual ASR models can also leverage cross-lingual information. Although one could perform a hyperparameter search over the number of output units, this could be computationally expensive and may not be feasible in practice.

Table~III compares the 1-hot and 2-hot WER for csvMASR-3. It can be seen that the WER is generally lower on the diagonal entries than the off-diagonals, indicating that the model performance can benefit from prompting. The gaps are also small ($<1\%$ WER), indicating that the model can automatically configure itself.

In addition, we also obtain the layer classification accuracy (LCA) for each adapters layer. Table~IV shows the LCA for each adapters layer in csvMASR-3 and Framewise-3. Although the accuracies are similar for the 2nd and 3rd layers, the difference is more pronounced for the 1st layer. In particular, csvMASR-3 has up to $16.65\%$ higher accuracy than Framewise-3. This indicates that our model's design tackles the language confusion issue better than Framewise-3. 

\begin{figure}[!t]
  
  \centering
  \vspace{-3mm}
  \includegraphics[width=\columnwidth]{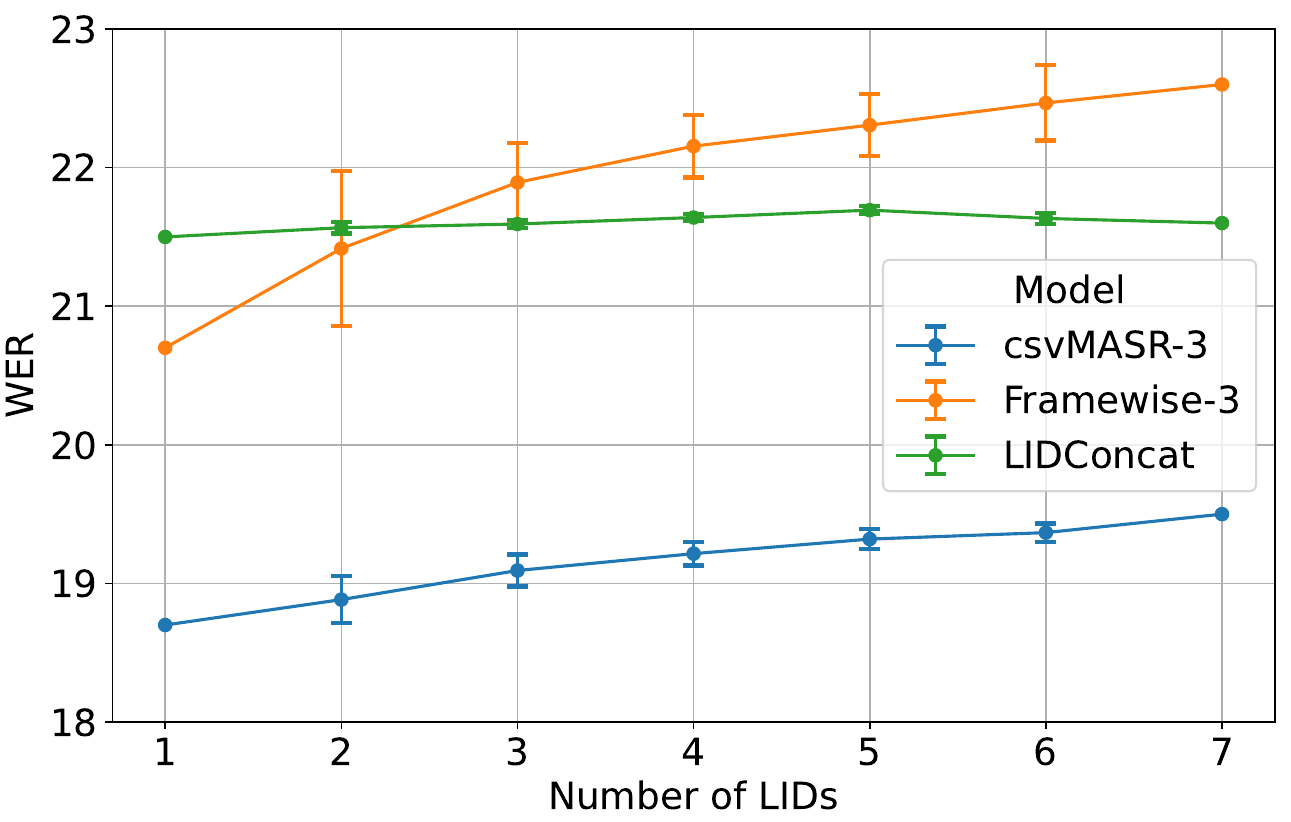}
  \vspace{-7mm}
  \caption{WER performance on the Portuguese test set by varying the number of additional LID vectors. All $2^6$ possible combinations of LID vectors are used and the mean and $1.96\times\text{standard error}$ (over the number of LID vectors) are plotted. NAR decoding is used for this illustration.}
  \label{fig:1:WER_growth}
  \vspace{-5mm}
\end{figure}

Lastly, we take Portuguese and show the change in WER over varying numbers of additional LID vectors in Figure~\ref{fig:1:WER_growth}. We can see that LIDConcat curve is mostly flat (from $18.7\%$ to $18.8\%$) and the Framewise-3 curve is steeper (from $20.7\%$ to $22.6\%$), confirming our previous observations of configurability. In contrast, the csvMASR-3 curve's slope is moderately flat in comparison (from $18.7\%$ to $19.5\%$). However, unlike LIDConcat, we mentioned previously that inference without ground-truth LID devastates the performance of csvMASR-3, indicating that the model relies heavily on the language-specific adapters. This indicates that csvMASR's flatness is due to its strong automatic configurability.

\begin{table}[!t]
\label{tab:1:csvMASR_fewhot_wer_scores}
\vspace{-2mm}
\caption{1-hot and 2-hot AR WER for csvMASR-3. }
\vspace{-4mm}
The diagonal entries are 1-hot WER and off-diagonals are 2-hot WER, where entry $(i, j)$ means WER on ground-truth $j$ with $(i,j)$ as the input LID.
\centering
\begin{adjustbox}{max width=\columnwidth}
 \begin{tabular}{|c|c|c|c|c|c|c|c|c|c|}
 \hline
 Language & Dutch & French & German & Spanish & Italian & Portuguese & Polish \\ \hline
 Dutch & \textbf{12.6} & 7.4 & 8.1 & 5.9 & 14.0 & 16.3 & 17.1 \\
 French  & 12.6 & \textbf{7.4} & 8.1 & 5.9 & 14.0 & 16.2 & 17.1 \\
 German & 12.6 & 7.5 & \textbf{8.1} & 5.9 & 14.0 & 16.3 & 17.2 \\
 Spanish & 12.6 & 7.5 & 8.1 & \textbf{5.9} & 14.2 & 16.5 & 17.2 \\
 Italian & 12.6 & 7.5 & 8.2 & 6.0 & \textbf{14.0} & 16.8 & 17.1 \\
 Portuguese & 12.6 & 7.5 & 8.1 & 5.9 & 13.9 & \textbf{16.3} & 17.3 \\
 Polish & 12.6 & 7.5 & 8.1 & 5.9 & 14.0 & 16.7 & \textbf{17.1} \\
 \hline
 \end{tabular}
\end{adjustbox}
\end{table}

\begin{table}[!t]
  \vspace{-4mm}
  \label{tab:1:classification_scores}
  \centering
  \caption{Language classification accuracies for each adapter layer. All-hot LID prompting is used.}
  \vspace{-3mm}
  \begin{adjustbox}{max width=\columnwidth}
    \begin{tabular}{|c|c|c|c|c|c|c|}
      \hline
  & \multicolumn{3}{c|}{Portuguese}& \multicolumn{3}{c|}{Polish} \\
      \hline
 Model / Adapters Layer & 1 & 2 & 3 & 1 & 2 & 3 \\
      \hline 
 csvMASR-3 & 69.35 & 97.24 & 99.08 & 69.23 & 99.62 & 100.00 \\
 Framewise-3 & 52.7 & 98.28 & 98.62 & 59.42 & 100.00 & 100.00 \\
      \hline
      \end{tabular}
\end{adjustbox}
\end{table}

 \section{Conclusion}
\label{sec:conclusion}
We introduce the Configurable MASR model with Summary Vector (csvMASR) model, a configurable MASR architecture. csvMASR leverages parameter-efficient adapters, a novel configurable weighted interpolation mechanism using speech summary representations and a language classification loss. We show that csvMASR can reduce the baseline WER from 10.33\% to 9.95\% on a 7-language setup. csvMASR also performs strongly in language classification with up to 16.65\% higher accuracy than the Framewise model. Lastly, language prompting tasks demonstrate its configurability, with a WER gap of $<1\%$ between 1-hot and all-hot LID inference. Given the parameter-efficiency of csvMASR, future work could include scaling it up to more languages.

\bibliographystyle{IEEEbib}
\bibliography{ref}

\end{document}